\documentstyle[11pt,paspconf]{article}

\begin{document}

\title{Line-Strength Gradients in Cooling Flow Galaxies}

\author{N. Cardiel and J. Gorgas}
\affil{Departamento de Astrof\'{\i}sica, Universidad Complutense de Madrid,
28040-Madrid, Spain}

\author{A. Arag\'{o}n-Salamanca}
\affil{Institute of Astronomy, Madingley Road, Cambridge CB3 OHA, UK}

\begin{abstract}
We present new results on line-strength gradients in a sample of 18
brightest cluster galaxies (13 in clusters with cooling flows and 5 in
clusters without). Here we focus on the study of the Mg$_2$ index and
the 4000\AA\ break (D$_{4000}$). We find that line-strength gradients
vary markedly from galaxy to galaxy, depending both on the mass
deposition rate and the presence or not of emission lines in the
nuclear regions.  Gradients are found to be flat, and even positive
(i.e. bluer when going inwards), in the emission-line region of the cooling
flow galaxies with emission lines. However, outside the region where
emission lines are present, mean spectral gradients of brightest
cluster galaxies, in clusters with and without cooling flows, are
consistent with those observed in giant elliptical galaxies.
In addition, and in agreement with previous studies, we confirm a
correlation between central spectral indices and the mass deposition
rates, although we find that cooling flow galaxies without emission
lines do not follow this trend. 
\end{abstract}


\section{Introduction}

The study of spectral absorption features provides a direct test of 
the possible presence of young stellar population in brightest cluster
galaxies (BCG's) immersed in cooling flow clusters (Sarazin 1986).
Cooling flows have been revealed in X-rays studies, but so far work in
the optical has failed to produce substantial {\it direct\/} evidence
for their presence (Fabian 1994).  The mechanism(s) responsible for the
connection, if any, between the emission line regions usually found at
the centers of many cooling flow galaxies, the blue excess, and the
cooling flow phenomenon itself remain unclear (Baum 1992, Fabian 1994,
and references therein).
%
Line-strength gradients are a powerful tool to investigate the radial
distribution of stellar populations in galaxies. They could be a
measurement of the amount of dissipation that occurred during the star
formation phase in the galaxies (see Gonz\'{a}lez \& Gorgas 1996 for a
recent review on spectroscopic gradients in early-type galaxies).
Typically, metallic line-strengths decrease outwards in elliptical
galaxies, which has conventionally been interpreted as a decrement of
me\-ta\-lli\-ci\-ty with radius (Faber 1977). However,  age effects may also
play an important role (Arag\'on, Gorgas \& Rego 1987; Gorgas, Efstathiou
\& Arag\'{o}n-Salamanca 1990; Gonz\'{a}lez 1993, Faber {\it et al.\/}
1995). The application of evolutionary population synthesis models
can help to disentangle this degeneracy between age and metallicity (Worthey
1994).
In this talk, we highlight results derived from an extension of the
work presented in Cardiel, Gorgas \& Arag\'{o}n-Salamanca
(1995) after the inclusion of extensive new spectroscopic data, obtained with
the WHT.

\section{Mg$_2$ and the 4000\AA\ break}

We have measured line-strength gradients in a sample of 18 galaxies (13
in clusters with cooling flows and 5 in clusters without). We
concentrate here on the study of the Mg$_2$ index and the
4000\AA\ break.  Both spectral features are good indicators of changes
in the stellar populations of early-type galaxies and can be measured
in spectra with relatively low signal-to-noise ratios.  This last
property is important when trying to obtain reliable measurements out
to $r \sim r_{\rm e}$, where the surface brightness of a typical BCG is
only a few per cent of the sky brightness.  In Figure~1 we show
4000\AA\ gradients for six galaxies in our sample.

\begin{figure}
\vspace{6.5in}
\caption{
D$_{4000}$ gradients for six galaxies in our sample
($H_0=75\,$km$\,$s$^{-1}\,$Mpc$^{-1}$ assumed). The thick horizontal
solid line (upper left) indicates the spatial extension of the emission
lines (when present). Mass deposition rates in $M_\odot/$yr (Edge {\it
et al.\/} 1992) are given in the upper right corner. Filled circles and
stars refer to different sides of the galaxies. Open symbols correspond
to secondary nuclei. Lines represent error-weighted least-squares fits,
excluding data with $r < 1.5\,$arcsec (affected by seeing) and
secondary nuclei.
}
\end{figure}

Mg$_2$ has been extensively studied by Gorgas {\it et al.\/} (1993) and
Worthey {\it et al.\/} (1994), who derived empirical fitting functions
to model the behavior of this spectral feature with the stellar
atmospheric parameters. D$_{4000}$ is quite sensitive to the
temperature of the main-sequence turn-off (Hamilton 1985, Dressler \&
Shectman 1987), although its dependence on metallicity is not
negligible (Worthey 1994). The corresponding fitting functions for this
feature are under construction (Gorgas \& Cardiel 1996). When these
empirical calibrations are incorporated into stellar population models
(Worthey 1994, Charlot \& Bruzual 1996), we will be able to interpret
the observed line-strengths in terms of mean age and metallicity of the
stellar populations.

It is important to note that the Balmer-line indices (H$\beta$,
H$\gamma_A$, H$\delta_A$, or H$\gamma_{\rm HR}$), widely used as 
age discriminators for stellar population in elliptical ga\-la\-xies
(Gonz\'{a}lez 1993, Rose 1994, Faber {\it et al.\/} 1995, Jones \&
Worthey 1995), are useless for such purpose here since, in many cases,
prominent emission prevents the observation of the underlying
absorption features in cooling flow galaxies. Moreover, in order to get 
reliable line indices, emission lines within the Mg$_2$ and
D$_{4000}$ bandpasses (e.g. [NeIII] $\lambda$3869, [SII]
$\lambda\lambda$4069,4076, H$\delta$, [OIII] $\lambda$4959, [NI]
$\lambda$5200) must be carefully removed (see details in Cardiel,
Gorgas \& Arag\'{o}n-Salamanca 1996) prior to any measurement.

\section{Results}
Full details of the new galaxy sample and measurements
will be given in a forthcoming paper (Cardiel, {\it et al.\/}
1996). We summarize here the main results which emerge from this study.
\begin{itemize}
  \item Central\footnote{Inside a fixed aperture of 4
arcsec projected at the distance of the Coma cluster, i.e. 1.8$\,$kpc
for $H_0=75\,$km$\,$s$^{-1}\,$Mpc$^{-1}$.}
Mg$_2$ and D$_{4000}$ measurements in cooling flow
galaxies {\it with emission lines\/}   
are well correlated with mass deposition rate, in the
sense that both indices decrease ---suggesting the presence of a young
stellar population--- when the mass flow increases.  Similar
correlations of central blue colors and spectral features with mass
rates have been previously reported by different authors:  Johnstone,
Fabian \& Nulsen (1987) found a correlation between the D$_{4000}$ and
the mass flow rate $\dot{M_{\rm V}}$ within their spectrograph slit;
McNamara \& O'Connell (1989)  presented a correlation between
their $\delta$UB (which is well correlated with D$_{4000}$) and the
normalized rate of mass drop out in the nucleus per unit luminosity
$(\beta/\beta_0)_{\rm nuc}$; finally, McNamara \& O'Connell (1992) showed
that the strength and extend of bluer central colors are more prominent
for central galaxies in clusters with high mass accretion rates.
  \item Interestingly, nuclear indices in the cooling flow galaxies of
our sample {\it without\/} emission lines (e.g. A~644, A~2142) {\it do not\/} 
follow this correlation with $\dot{M}$, which is also the case for
A~2029 in the $\delta$UB--$(\beta/\beta_0)_{\rm nuc}$ diagram from
McNamara \& O'Connell (1989). The central line-strengths of BCG's in both 
clusters without cooling flows and cooling flow galaxies without
emission lines are consistent with the values observed in giant
ellipticals.
  \item The central line-strengths of elliptical galaxies define a
relatively narrow trend in the D$_{4000}$--Mg$_2$ plane (Kimble,
Davidsen \& Sandage 1989). However, central Mg$_2$ and D$_{4000}$
measurements in cooling flow galaxies exhibit a clear correlation (see
Figure~5 in Cardiel {\it et al.\/} 1995) which departs from the locus
where giant elliptical galaxies are found.  The inclusion of new
galaxies in our sample has revealed that there is a clear dichotomy in
the way cooling flow galaxies with emission lines populate this plane.
Galaxies with central emission-line nebulae of class II (following the
scheme of Heckman {\it et al.\/} 1989), e.g. PKS~9745$-$191, Hydra~A,
A~1795, A~2597, exhibit the largest departures from the elliptical
galaxy region in the D$_{4000}$--Mg$_2$ diagram. When compared with
stellar population model predictions this provides strong evidence for
recent star formation in these galaxies.
  \item Although line-strength gradients are usually found to be linear 
with $\log(r)$ in elliptical galaxies (Gonz\'{a}lez \& Gorgas 1996),
Mg$_2$ and D$_{4000}$ gradients in some cooling flow galaxies exhibit a
clear slope change at intermediate radii (Cardiel {\it et al.\/} 1995).
With the inclusion of new spectroscopic data, we find that this change
of slope is only evident for galaxies with emission lines (Figure~1).
  \item Cooling flow galaxies {\it with emission lines} exhibit flat 
and even positive gradients in the region where the emission is
detected. In fact, these inner gradients seem to be correlated with
mass deposition rate (i.e., more positive with increasing $\dot{M}$;
see Figure~2).  In the outer parts of these galaxies, where emission is
not observed, the derived mean line-strength gradients are consistent
with those in elliptical galaxies.
  \item Mean gradients in galaxies with cooling flows but {\it without
emission lines} are similar to those in galaxies without cooling flows
and those in elliptical galaxies (Figure~3).  \end{itemize}

\begin{figure}
\vspace{3.0in}
\caption{
D$_{4000}$ and Mg$_2$ gradients for the galaxies of our sample versus
mass deposition rate (in $M_\odot$/yr).
Open symbols correspond to inner gradients (i.e. in the
emission line region) of cooling flow galaxies, whilst closed circles
stand for galaxies without cooling flows and cooling flow galaxies
without emission lines.
}
\vspace{2.9in}
\caption{
Comparison of mean D$_{4000}$ and Mg$_2$ gradients for normal
ellipticals (E's), brightest cluster galaxies without cooling flows (BCG's No
CF), cooling flow galaxies without emission lines (CFG's No EL), and cooling 
flow galaxies with emission lines (CFG's EL) in the inner (i.e. emission 
line) and outer (out from the emission line) regions. Error bars indicate
the formal errors in the mean values.
}
\end{figure}

\section{Discussion}

$\bullet$ {\bf Blue excess and emission lines.} Previous work has shown
that in the central parts of massive cooling flows the emission line
luminosity correlates with the blue light excess (Johnstone {\it et
al.} 1987, Allen {\it et al.} 1992, Crawford \& Fabian 1993, Crawford
{\it et al.} 1995).  In addition, the emission line flux is co-spatial
with the excess blue light (e.g.  Romanishin 1987, Hansen, J$\o$rgensen
\& N$\o$rgaard-Nielsen 1995).  Our Mg$_2$ and D$_{4000}$ gradients
confirm that the spatial extent of the excess blue light is very
similar to the emission line region (Figure~1).  These results indicate
that, very likely, both processes are physically related.  However this
is a question of open debate (see discussion in Baum 1992, Fabian 1994,
and references therein).  Using blue optical spectra and IUE data,
Crawford \& Fabian (1993) showed that the spectrum of the excess blue
light is consistent with both star formation and a power-law spectrum.
Allen (1995), using simple stellar population models, suggested that
young stars are responsible of the spatially extended excess UV/blue
continua. Our analysis of central Mg$_2$ and D$_{4000}$ measurements in
a sample of BCG's with and without cooling flows using Bruzual \&
Charlot (1993) stellar population models suggests the observed
correlation between both indices can be reproduced with star formation
at a constant rate over the last few Gyrs (Cardiel {\it et al.\/}
1995). However, further work in the stellar population modelling is
still needed to definitely rule out a power-law spectrum as the origin
of the blue excess.


\noindent $\bullet$ {\bf The cooling flow connection.} The presence of
emission lines must be in some way related to cooling flows since
H$\beta$ (Johnstone {\it et al.\/} 1987) and H$\alpha$ luminosity
(Heckman {\it et al.\/} 1989) are correlated with mass deposition rate.
However, considering the high scatter in line luminosities for a given
$\dot{M}$, the relationship does not seem to be a simple one (Baum 1992).
Whichever me\-cha\-nism is responsible for the dilution of the measured
line-strength indices (star formation, scattering of nuclear light,
etc.), the correlations of the mass deposition rate with central
indices and line-strength gradients in the emission line regions add
further weight to the idea of a link between the blue excess and the
cooling flow phenomenon. Nevertheless, as 
Crawford \& Fabian (1993) pointed out, the fact that there are central galaxies
(e.g. A~644, A~2029) in clusters with high $\dot{M}$ but without blue
excess suggests that the cooling flow alone cannot be responsible for
the blue light.  These authors also noted the existence of strong
cooling flow galaxies, with (e.g. A~478) and without (e.g. A~2029)
emission-line nebulae, which apparently did not exhibit a blue
continuum excess. However, our D$_{4000}$ gradient in the central
galaxy of A~478 (Figure 1) does exhibit a clear positive behaviour
(i.e. blue excess) in the emission line region. 
A larger sample is needed to test whether there are any cooling flow galaxies
with line-emission nebulae which do not exhibit blue excess, and vice versa.

\acknowledgments

We thank the organizing committee for making possible this meeting.
The WHT is operated on the island of La Palma by the Royal Greenwich
Observatory at the Observatorio del Roque de los Muchachos of the Instituto
de Astrof\'{\i}sica de Canarias.
This work was supported in part by the Spanish `Programa Sectorial de
Promoci\'{o}n General del Conocimiento' under grant No. PB93-456. AAS
acknowledges generous financial support from the Royal Society. 


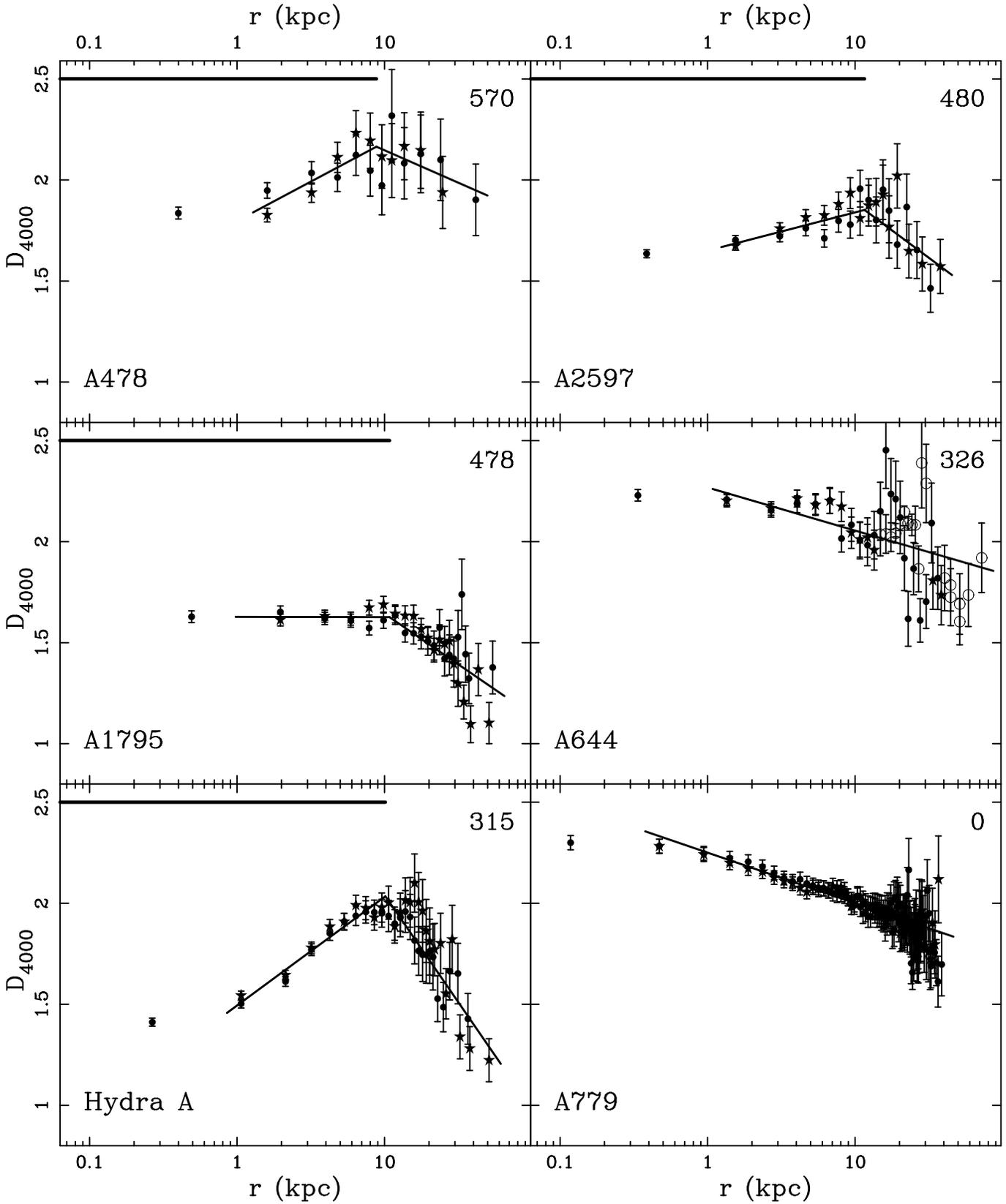
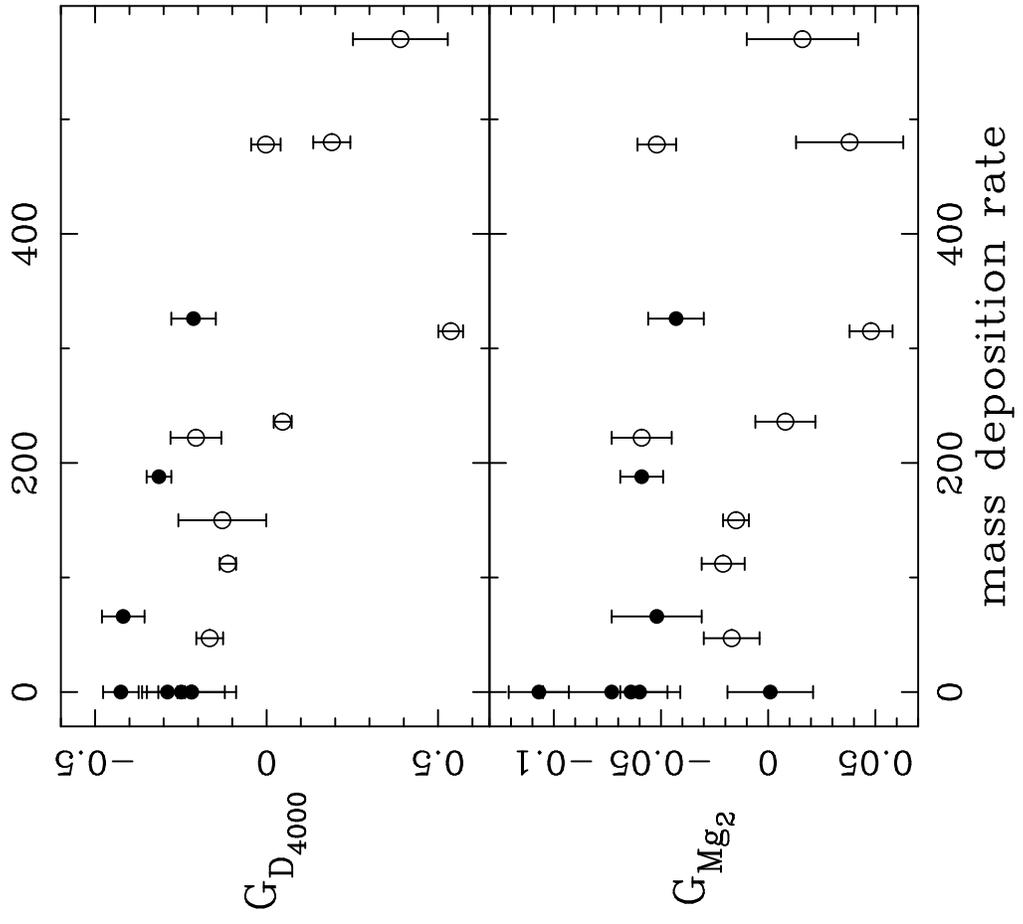
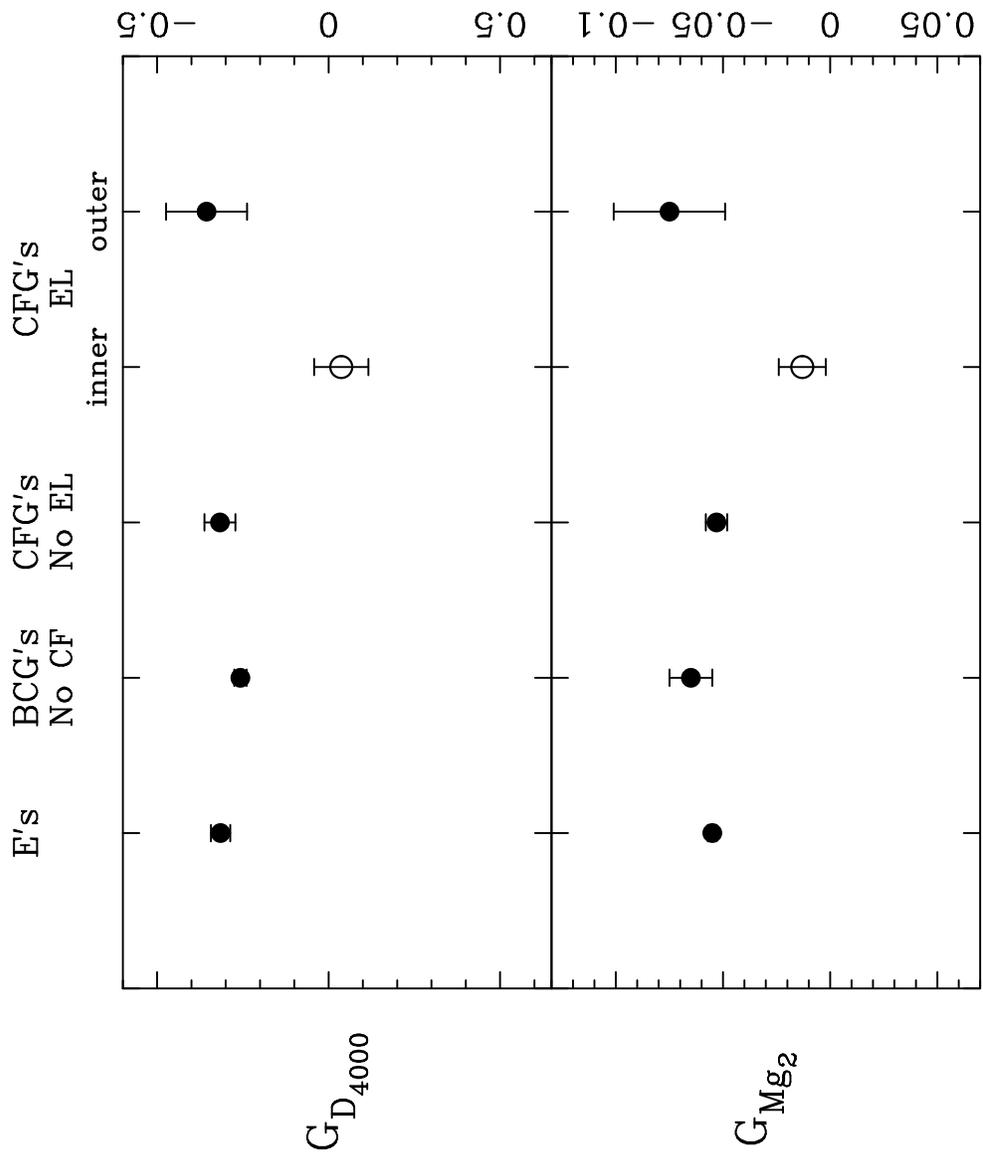
\end{document}